\documentclass[traditabstract,twocolumn]{aa} % for a referee version
\usepackage{graphicx}
\usepackage{txfonts}
\usepackage{natbib}
\usepackage{subfigure}
\newcommand{\tablenotea}[1]{\parbox{8.8cm}{ \indent
\footnotesize{\textsc{Note.--}~#1}}}
\newcommand{\nature}{Nature}                      % Nature
\newcommand{\jms}{J.~Mol.~Spectr.}                %
\newcommand{\sacta}{Spectrochim.~Acta~Part~A}     %

\begin{document}
\title{\emph{Herschel}/HIFI\thanks{Herschel is an ESA space observatory with science instruments
provided by European-led Principal Investigator consortia and with
important participation from NASA.} observation of highly excited
rotational lines of HNC toward IRC~+10\,216\thanks{Based on
observations carried out with the IRAM 30-meter telescope. IRAM is
supported by INSU/CNRS (France), MPG (Germany) and IGN (Spain).}}
\titlerunning{Highly excited HNC in IRC~+10\,216}
\authorrunning{Daniel et al.}
\author{F. Daniel\inst{1}, M. Ag\'undez\inst{2}, J. Cernicharo\inst{1}, E. De Beck\inst{3}, R. Lombaert\inst{3}, L. Decin\inst{3,4}, C. Kahane\inst{5}, M. Gu\'elin\inst{6}, H. S. P. M\"{u}ller\inst{7}}
\institute{Departamento de Astrof\'isica, Centro de
Astrobiolog\'ia, CSIC-INTA, Ctra. de Torrej\'on a Ajalvir km 4,
28850 Madrid, Spain; \email{danielf@cab.inta-csic.es} \and LUTH,
Observatoire de Paris-Meudon, 5 Place Jules Janssen, 92190 Meudon,
France \and Instituut voor Sterrenkunde, Katholieke Universiteit
Leuven, Celestijnenlaan 200D, 3001 Leuven, Belgium \and
Sterrenkundig Instituut Anton Pannekoek, University of Amsterdam,
Science Park 904, NL-1098, Amsterdam, The Netherlands \and
Laboratoire d'Astrophysique de l'Observatoire de Grenoble, 38041
Grenoble, France \and Institut de Radioastronomie Millim\'etrique,
300 rue de la Piscine, 38406 Saint Martin d'H\`eres, France \and
I. Physikalisches Institut, Universit\"at zu K\"{o}ln,
Z\"{u}lpicher Str. 77, 50937 K\"{o}ln, Germany}

\date{Received; accepted}

% \abstract{}{}{}{}{}
% 5 {} token are mandatory

\abstract
{We report the detection in emission of various highly excited
rotational transitions of HNC ($J$ = 6-5 through $J$ =12-11)
toward the carbon-star envelope IRC~+10\,216 using the HIFI
instrument on-board the Herschel Space Observatory. Observations of
the $J$ = 1-0 and $J$ = 3-2 lines of HNC with the IRAM 30-m
telescope are also presented. The lines observed with HIFI have
upper level energies corresponding to temperatures between 90 and
340 degrees Kelvin, and trace a warm and smaller circumstellar
region than that seen in the interferometric maps of the $J$ = 1-0
transition, whose emission extends up to a radius of 20\arcsec. After
a detailed chemical and radiative transfer modeling, we find that
the presence of HNC in the circumstellar envelope of IRC~+10\,216 is
consistent with formation from the precursor ion HCNH$^+$, which
in turn is produced through several proton transfer reactions
which are triggered by the cosmic-ray ionization. We also find
that the radiative pumping through $\lambda$ 21 $\mu$m photons to
the first excited state of the bending mode $\nu_2$ plays a
crucial role to populate the high-$J$ HNC levels involved in the
transitions observed with HIFI. Emission in these high-$J$
rotational transitions of HNC is expected to be strong in regions
which are warm and dense and/or have an intense infrared flux at
wavelengths around 21 $\mu$m.}


\maketitle
%
%________________________________________________________________

\section{Introduction}

Hydrogen isocyanide (HNC), a metastable isomer of HCN lying 0.6 eV
higher in energy, is ubiquitous in the interstellar
medium. First observed in several objects through a line at
90.7 GHz (e.g. \citealt{zuc72}), the definitive confirmation of
this line, the $J$ = 1-0 rotational transition of
HNC, came through spectroscopic laboratory experiments in the late
1970s \citep{bla76}. HNC is widely observed in different types of
astronomical regions, such as diffuse clouds \citep{lis01}, cold
dark clouds \citep{hir98}, star-forming regions \citep{sch92},
circumstellar envelopes around evolved stars \citep{buj94},
external galaxies \citep{aal02}, and even in high redshift objects
\citep{gue07}. The main formation mechanism of HNC is thought to
be the dissociative recombination of the ion HCNH$^+$
\citep{sem01}. HNC has mostly been observed only in the $J$ = 1-0
transition at $\sim$3.3 mm, preventing to carry out a detailed excitation analysis \citep{sar10}.

The HIFI instrument on-board the Herschel Space Observatory,
operating in the 480-1910 GHz frequency range, allows to cover
high-energy rotational transitions of HNC. In this study, we
present the detection in the C-star envelope of IRC~+10\,216 of the $J$
= 6-5 through $J$ = 12-11 rotational transitions of HNC.
Observations of the $J$ = 1-0 and $J$ = 3-2 transitions obtained
with the IRAM 30-m telescope are also presented. The large set of
observed lines allows us to investigate in detail the abundance
distribution and excitation of HNC in IRC~+10\,216's circumstellar envelope.

\section{Observations and results}

\begin{figure}
\begin{center}
\includegraphics[angle=0,scale=.56]{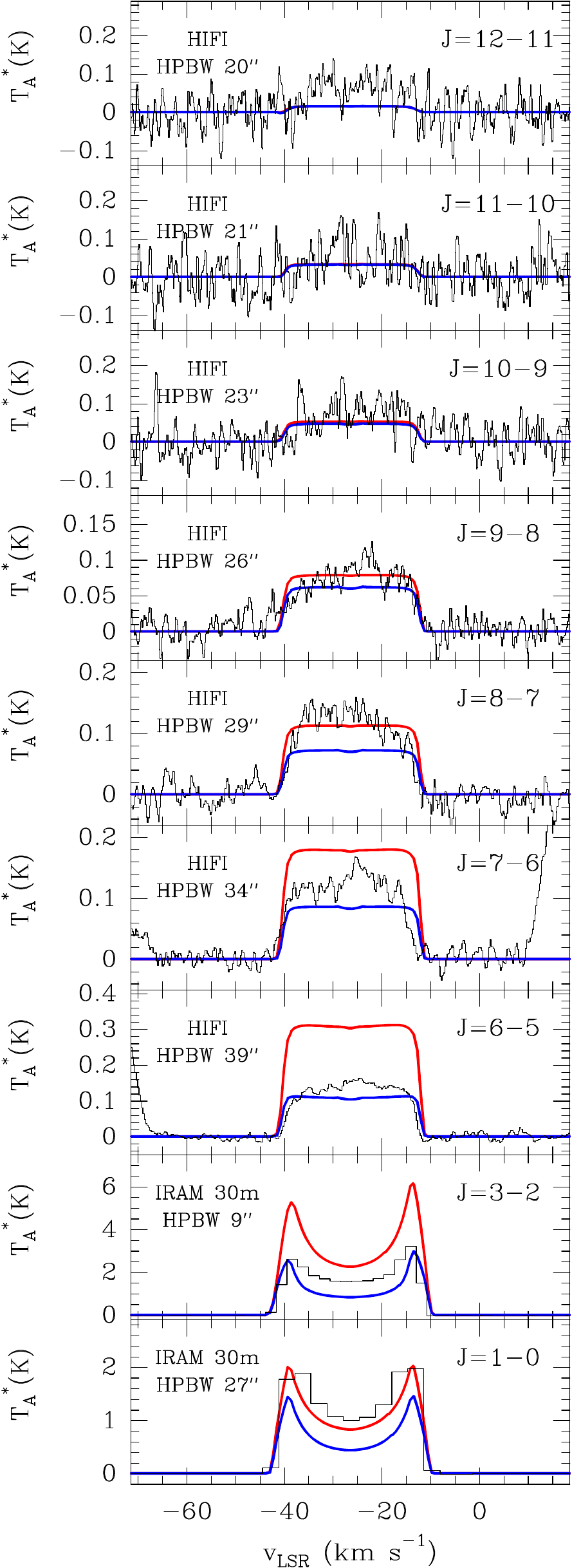}
\caption{HNC rotational lines observed toward IRC~+10\,216 with the
IRAM 30-m telescope ($J$ = 1-0 and $J$ = 3-2) and with HIFI 
($J=6-5$ to $J=12-11$). The red and blue curves correspond to the models
described in Sect. \ref{sec:model} and \ref{sec:results}.} \label{fig-hnc-lines}
\vspace{-0.5cm}
\end{center}
\end{figure}

The HIFI observations were obtained in the context of a high
resolution line survey of IRC~+10\,216 \citep{cer10}. The data were
taken in double beam-switching mode with a spectral resolution of
1.1 MHz and channel spacings of 0.5 MHz, and processed using the
standard Herschel pipeline up to Level 2, providing fully
calibrated spectra. The HNC lines where all observed around the maximum
luminosity, i.e. phase $\phi = 0$. For details regarding the reduction procedure
we refer to \citet{cer10}. The total integration time per
frequency setting ranged from 2 to 14 minutes, resulting in
antenna temperature rms noise levels of 8-15 mK per 1.1 MHz
channel for the HNC spectra in bands 1a, 2a, 2b, and 3a, and of
50 mK for the spectra above 900 GHz. At the frequencies of the
HNC lines observed by HIFI ($J$ = 6-5 through $J$ = 12-11), the
beam sizes range from 39\arcsec\/ down to 20\arcsec\/, and is therefore comparable to
the size of the emitting region of HNC in IRC~+10\,216. The $J$ = 1-0 transition,
as mapped with the IRAM Plateau de Bure Interferometer (PdBI),
shows a ring-like distribution extending up to a radius of nearly
20\arcsec (see \citealt{gue97}), and the higher-$J$ transitions
observed with HIFI are expected to have a more compact
distribution.

The $J$ = 1-0 transition of HNC was observed before 2009 with the
IRAM 30-m telescope in the course of a $\lambda$ 3 mm spectral line
survey of IRC~+10\,216 (Cernicharo et al., in prep.). The
achieved $T_A^*$ rms noise level after averaging different spectra
at the frequency of the $J$ = 1-0 line is 3 mK per 1 MHz
channel. The $J$ = 3-2 line was observed in January 2010 during a
$\lambda$ 0.9 mm spectral line survey of IRC~+10\,216 (Kahane et al.,
in prep.), using the new EMIR receivers. The line was
observed with a spectral resolution of 2 MHz during about 3 h of
good atmospheric conditions ($T_{\rm sys}$ = 800 K) resulting in a
$T_A^*$ rms noise level of 6 mK per channel.
Additionnaly, we make use of the PdBI observations of the $J=1-0$ line. 
These observations are described in \cite{gue97} and references therein.

\begin{table}
\caption{HNC observed line parameters}
\label{table-lineparameters} \centering
\begin{tabular}{crllcc}
\hline \hline
\multicolumn{1}{c}{}           & \multicolumn{1}{c}{$\nu_{\rm 0}$} & \multicolumn{1}{c}{v$_{\rm LSR}$} & \multicolumn{1}{c}{v$_{\rm exp}$$^a$} & \multicolumn{1}{c}{$\int$$T_A^*$$d$v} & \multicolumn{1}{c}{$\eta$$^b$}  \\
\multicolumn{1}{c}{$J'-J''$}   & \multicolumn{1}{c}{(MHz)}         & \multicolumn{1}{c}{(km s$^{-1}$)}           & \multicolumn{1}{c}{(km s$^{-1}$)} & \multicolumn{1}{c}{(K km s$^{-1}$)} & \multicolumn{1}{c}{} \\
\hline
1-0   &   90663.6 & -26.3(1)  & 15.4(2)  & 44.9(8) & 0.80 \\ % january 2001 phase almost 0
3-2   &  271981.1 & -26.4(1)  & 14.4(1)  & 60.7(10)& 0.52 \\ % january 2010 phase almost 0
\hline
6-5   &  543897.6 & -26.2(3)  & 12.6(3)  & 3.53(5) & 0.75 \\
7-6   &  634510.8 & -26.7(4)  & 12.3(4)  & 2.75(6) & 0.75 \\
8-7   &  725107.3 & -26.7(4)  & 12.5(4)  & 2.87(8) & 0.75 \\
9-8   &  815684.7 & -25.6(7)  & 12.4(7)  & 1.95(15)& 0.75 \\
10-9  &  906240.5 & -25.8(15) & 10.5(20) & 1.23(25)& 0.74 \\
11-10 &  996772.3 & -27.4(15) & 13.6(18) & 1.55(22)& 0.74 \\
12-11 & 1087277.9 & -25.3(10) & 12.2(12) & 0.94(20)& 0.74 \\
\hline
\end{tabular}
\tablenotea{Number in parentheses are 1$\sigma$ uncertainties in
units of the last digits. $^a$ v$_{\rm exp}$ is the half width at
zero level. $^b$ $\eta$ is the efficiency parameter used to
convert antenna temperature ($T_A^*$) into main beam brightness
temperature.}
\end{table}

The observed lines of HNC are shown in Fig.~\ref{fig-hnc-lines}
and the line and telescope parameters are given in
Table~\ref{table-lineparameters}. All the lines appear free of
contamination from other lines. The $J$ = 1-0 and $J$ = 3-2
lines observed with the IRAM 30-m telescope are the most intense
ones and show a U-shaped profile, typical of optically thin
emission resolved by the telescope. On the other hand, the lines
observed with HIFI show a flat-topped profile, which is typical of
optically thin lines not resolved by the telescope, i.e.
geometrically diluted in the main beam of HIFI.

\section{Modeling} \label{sec:model}

To learn about the excitation conditions and the formation
mechanism of HNC in the envelope of IRC~+10\,216 we have carried out
modeling tasks. The model, basically taken from \citet{fon08} and \citet{agu09},
considers the presence of a central star with an effective
temperature $T_*=2330$ K and a radius $R_*=4 \times 10^{13}$ cm,
surrounded by a spherical envelope of gas and dust. The adopted
mass loss rate is $2 \times 10^{-5}$ M$\odot$ yr$^{-1}$, with a
distance of 130 pc, and a terminal expansion velocity of 14.5 km
s$^{-1}$. Following \citet{cor09}, we consider the presence of
density-enhanced shells.

\begin{figure}
\begin{center}
\includegraphics[angle=270,scale=.36]{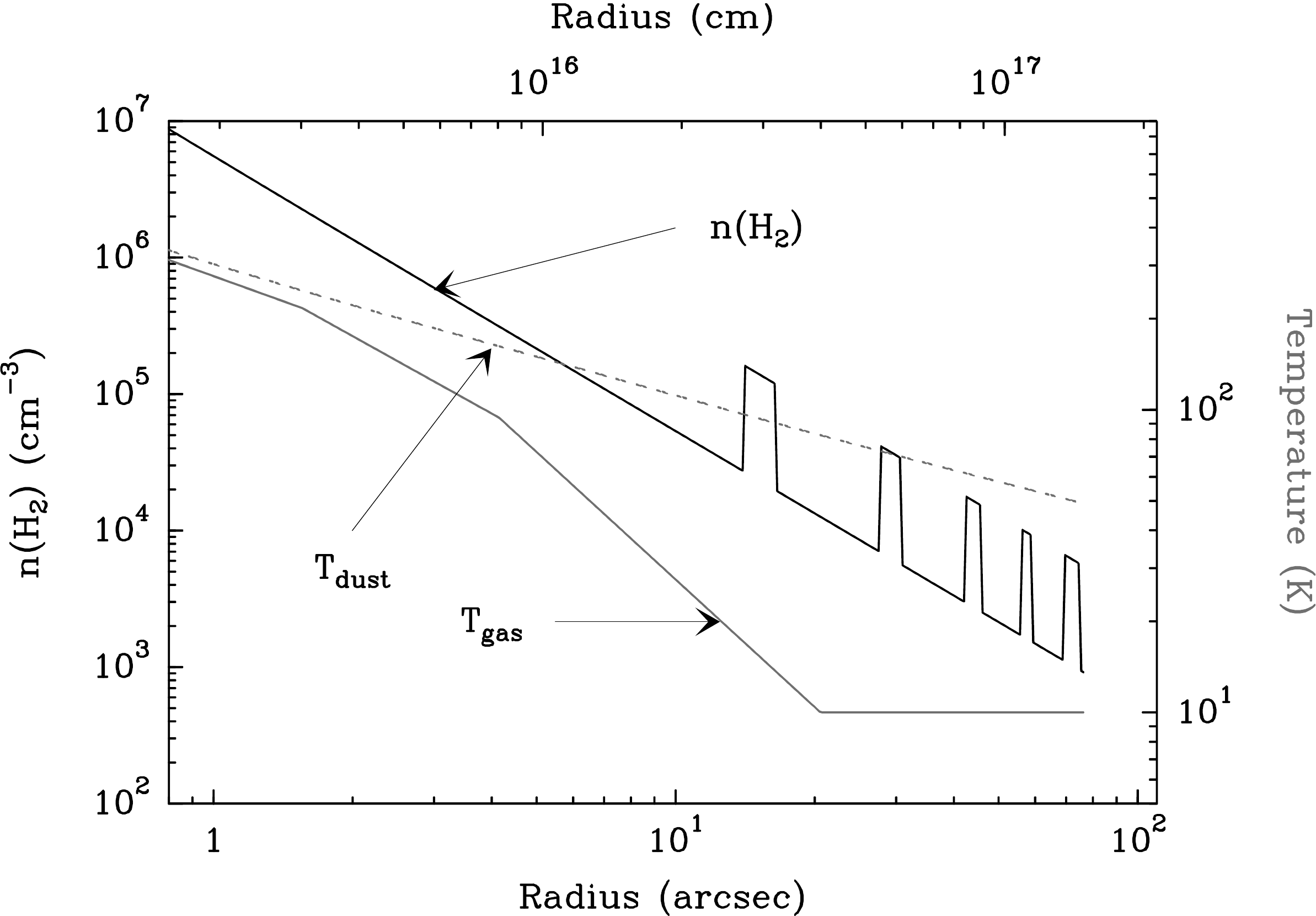}
\caption{Physical parameters (i.e. dust and gas temperatures, 
H$_2$ volume density) used in the present work to describe the circumstellar envelope.
density-enhanced shells are introduced in the current modeling as can be seen in the 
H$_2$ volume density curve. The first shell is located at 15\arcsec\/ and we assume an intershell distance
of 12\arcsec, the shells being 2\arcsec\/ wide.}
\label{fig10} \vspace{-0.5cm}
\end{center}
\end{figure}

The gas temperature profile was derived using $^{12}$CO and
$^{13}$CO HIFI and IRAM 30-m observations. The adopted $^{12}$CO
abundance relative to H$_2$ was $6 \times 10^{-4}$ and the
$^{12}$CO/$^{13}$CO abundance ratio 45 \citep{kah00}. The best fit
to the whole set of lines is given by a 3-fold power law, i.e.
$T_k \propto (r/R_*)^{-n}$ with $n=0.55$ for $r < 75 \, R_*$,
$n=0.85$ for $r < 200 \, R_*$ and $n=1.4$ for greater radii, with
the scale fixed by setting $T_k = T_*$ at $r = R_*$. The
estimated uncertainty in $T_k$ is of 50 \%.  The physical parameters 
that describe the circumstellar envelope, i.e. the dust and gas 
temperature as well as the H$_2$ volume density are presented in Fig. 
\ref{fig10}.

\begin{figure}
\begin{center}
\includegraphics[angle=-90,scale=.36]{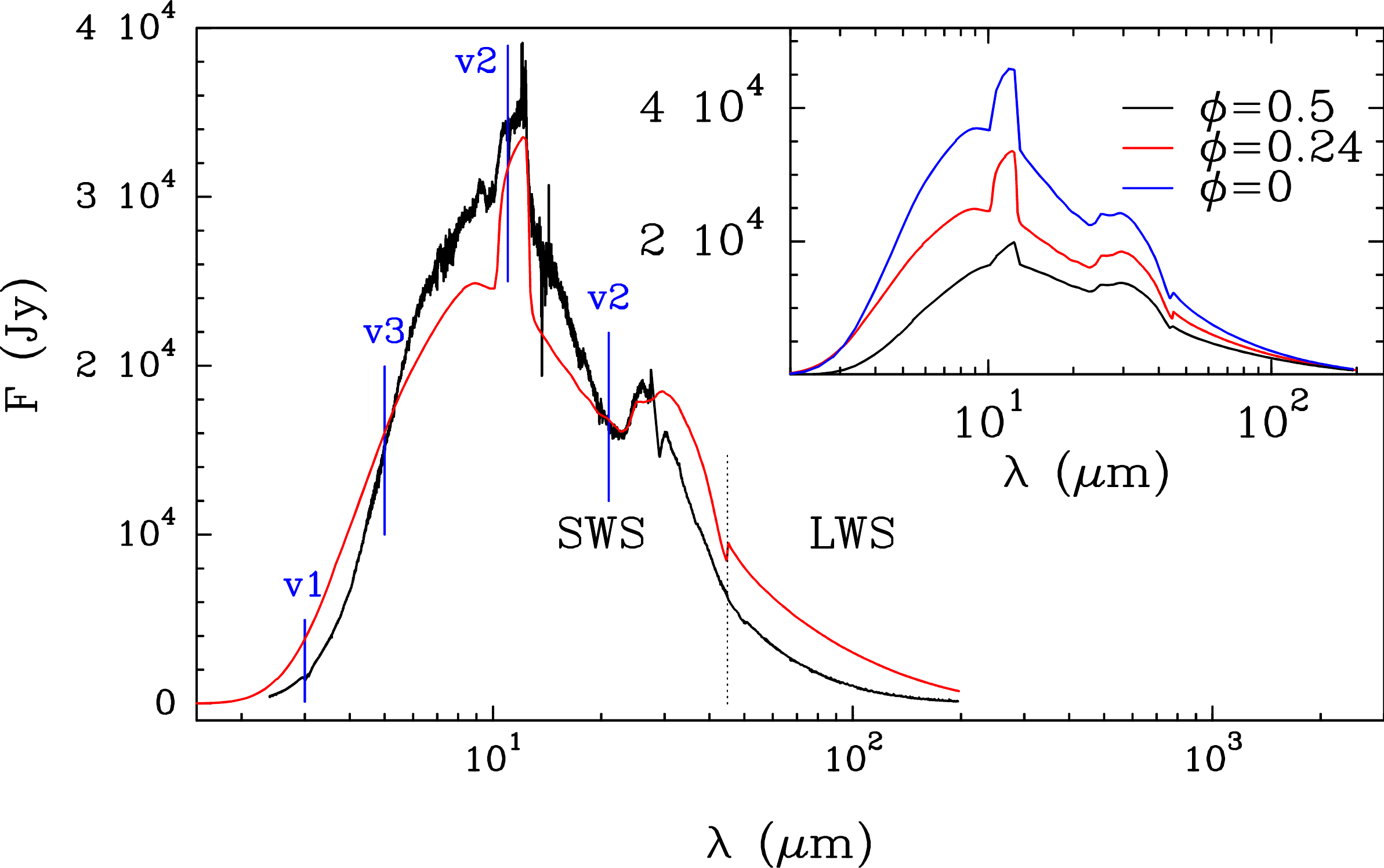}
\caption{Comparison between the SED of IRC~+10\,216 as modeled and
as observed by ISO at $\phi$ = 0.24. Wavelengths of some
vibrational transitions of HNC are indicated by vertical lines. 
The vibrational transition indicated at $\sim$ 11 $\mu$m correspond to the 
000 $\to$ 020 vibrational band and the transition at $\sim$ 21 $\mu$m
to the 000 $\to$ 010 and 010 $\to$ 020 bands.
The dotted vertical
line around 45 $\mu$m indicates the ISO LWS / SWS division.
The inset panel shows the computed SED at different phases.}
\label{SED} \vspace{-0.5cm}
\end{center}
\end{figure}

As will be shown below, infrared (IR) pumping to excited
vibrational states of HNC plays an important role in the
excitation of the observed lines. Hence, a good description of the
IR flux is needed. For that purpose the spectral energy
distribution (SED) observed by ISO at phase $\phi = 0.24$
\citep{cer99} has been modeled to derive dust opacities suitable
for IRC~+10\,216, which are given by a mixture of silicon carbide,
amorphous carbon and MgS dust grains (see Fig. \ref{SED}). The HNC
lines were observed at phase $\phi = 0.95-0.05$, around the
maximum luminosity. Therefore, the dust temperature and SED were recomputed modifying the star radius accordingly and using
photometric points observed at maximum \citep{leb92}. More details
concerning the modeling of IRC~+10\,216's SED can be found in \cite{debeck11}.
The parameters used in the present study to describe the dust (i.e. the dust opacities and dust-temperature profile) are the same as
those described in the latter article. The differences in the computed SEDs mainly come from the inclusion of the density-enhanced
shells (see Sect. \ref{sec:results}), of which the main effect is that the flux at wavelengths longer than 25 um is slightly increased.

\subsection{Chemical model} \label{section:chemistry}

The chemical composition of the gas has been computed as it
expands from a radius of $2 \times 10^{14}$ cm up to 10$^{18}$ cm.
To deal with the density-enhanced shells we have proceeded as
\citet{cor09}, running a model for the smooth component of the
envelope and another for a density-enhanced shell. The cosmic-ray
ionization rate of H$_2$ is taken as 1.2 $\times$ 10$^{-17}$
s$^{-1}$ \citep{agu06} and the ultraviolet interstellar radiation field of
\citet{dra78} is adopted. The chemical network comprises 470 gas
phase species (composed of H, He, C, N, O, Si, P, and S) linked by
7400 reactions, whose rate constants are taken from databases such
as udfa06\footnote{See \texttt{http://www.udfa.net}} \citep{woo07}
and from the recent literature on chemical kinetics. The
abundances at the initial radius are taken from \cite{agu09}.

\begin{figure}
\begin{center}
\includegraphics[angle=-90,scale=.42]{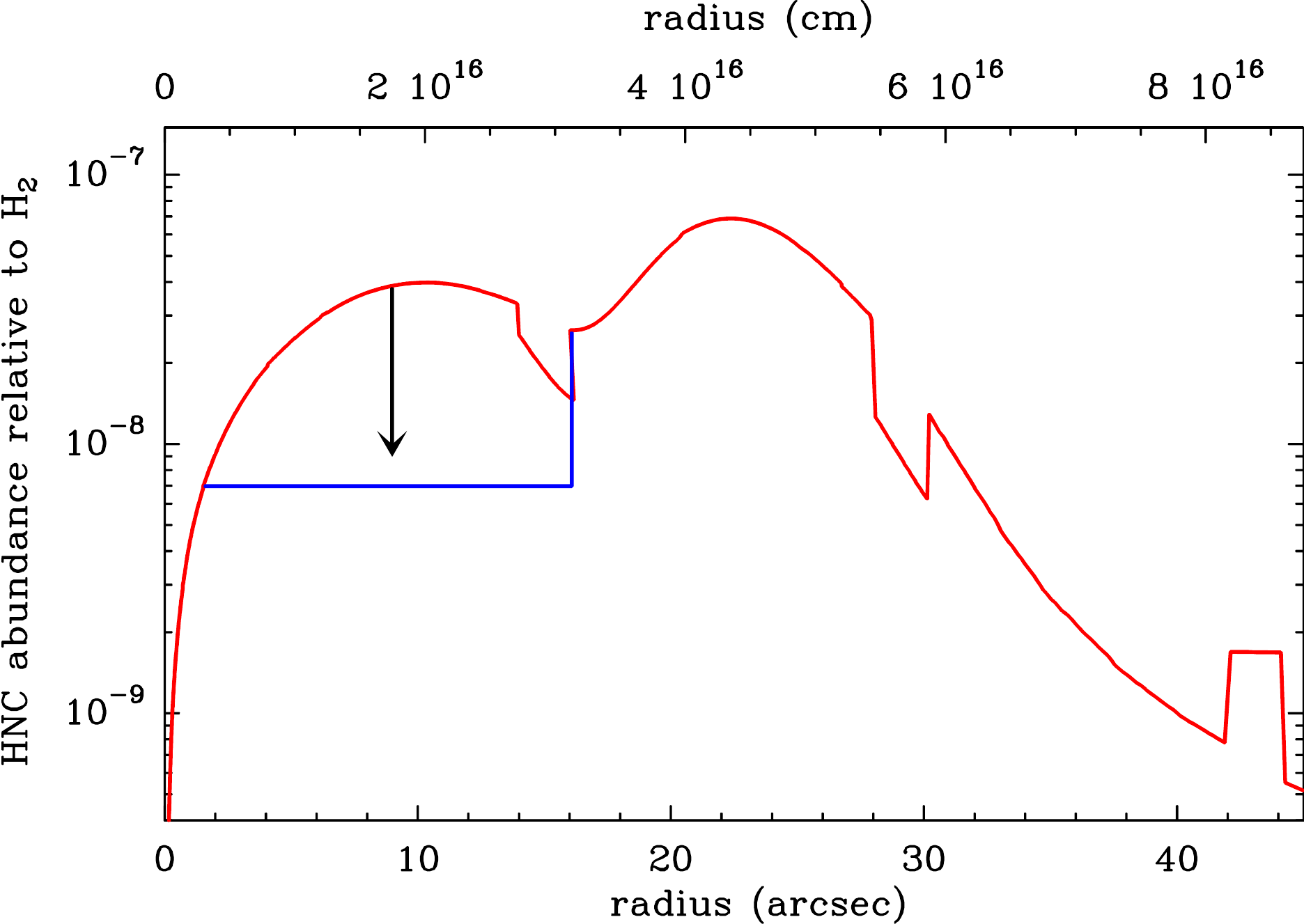}
\caption{Abundance distribution of HNC obtained with the chemical
model (red curve). The modification introduced to reproduce the
observations (blue lines in Fig.\ref{fig-hnc-lines} ) is shown by
to the blue curve.} \label{fig-hnc-abundance} \vspace{-0.5cm}
\end{center}
\end{figure}

\begin{figure}
\begin{center}
\includegraphics[angle=270,scale=.38]{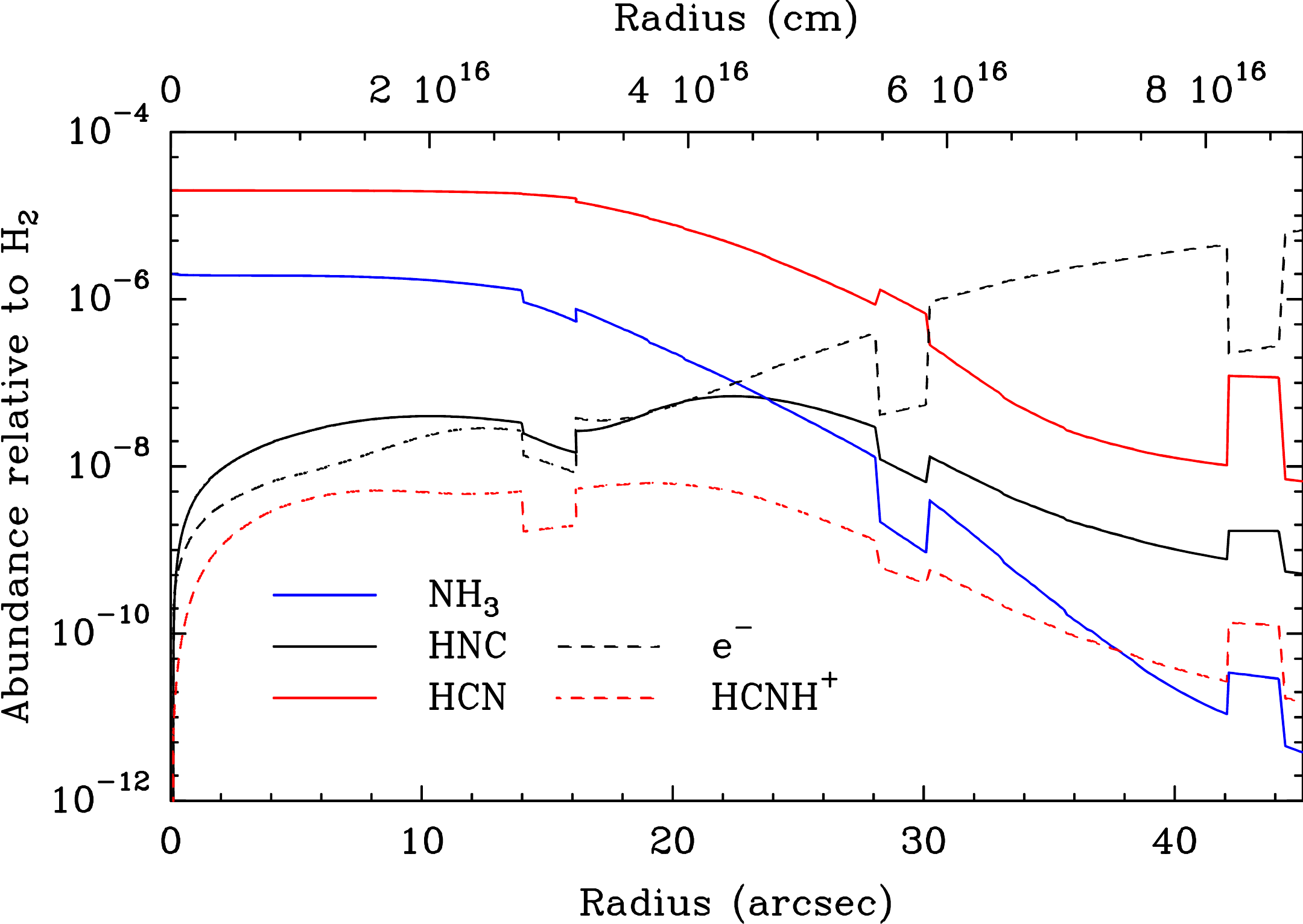}
\caption{Abundances of some of the species (i.e. HCNH$^+$, HCN, NH$_3$ and e-) related
to the HNC formation as obtained from the standard chemistry model. }
\label{fig11} \vspace{-0.5cm}
\end{center}
\end{figure}

The resulting abundance distribution of HNC in the envelope of IRC
+10216 is plotted in Fig.~\ref{fig-hnc-abundance} and the 
abundance of the species chemically related to HNC in Fig. \ref{fig11}. 
Apart from the
abrupt changes at 15\arcsec\/, 29\arcsec\/, and 43\arcsec\/, which correspond to the
presence of density-enhanced shells, we see that the fractional
abundance of HNC starts to increase at small radii, well before
the molecular shell located at 15\arcsec\/. This occurs because the
formation of HNC is, to a large extent, driven by cosmic rays that
can penetrate deeper in the envelope than interstellar ultraviolet
photons. In the regions inward of $10^{16}$ cm, HNC is mostly formed
through the reaction:
\begin{equation}
\rm HCNH^+ + NH_3 \rightarrow HNC + NH_4^+ \label{reac-hcnh+_nh3}
\end{equation}
Ammonia is relatively abundant in the inner layers of IRC~+10\,216
(2 $\times$ 10$^{-6}$ relative to H$_2$; \citealt{has06}), while
the precursor ion HCNH$^+$ is formed from several reactions of
proton transfer to HCN (a species which is also fairly abundant in
the inner envelope of IRC~+10\,216, $2 \times 10^{-5}$ relative to
H$_2$ according to \citealt{fon08}):
\begin{equation}
\rm HCO^+ + HCN \rightarrow HCNH^+ + CO \label{reac-hco+_hcn}
\end{equation}
\begin{equation}
\rm C_2H_3^+ + HCN \rightarrow HCNH^+ + C_2H_2
\label{reac-c2h3+_hcn}
\end{equation}
\begin{equation}
\rm H_3^+ + HCN \rightarrow HCNH^+ + H_2 \label{reac-h3+_hcn}
\end{equation}
These processes are driven by the cosmic-ray ionization of H$_2$
and the subsequent fast formation of H$_3^+$, which yields HCO$^+$
and C$_2$H$_3^+$ through proton transfer to CO and C$_2$H$_2$,
respectively. Reactions (\ref{reac-hcnh+_nh3}--\ref{reac-h3+_hcn})
have been experimentally found to be rapid (rate constants are in
excess of 10$^{-9}$ cm$^3$ s$^{-1}$ at 300 K; \citealt{ani03}).
For reaction (\ref{reac-hcnh+_nh3}), however, it is not clear what
are the branching ratios yielding HCN and HNC, so that we
consider that both channels occur at the same rate, as is done by e.g. the databases udfa06$^1$ and KIDA\footnote{See
\texttt{http://kida.obs.u-bordeaux1.fr}}. This assumption will be discussed in Sect. \ref{sec:results}.

The peak abundance of HNC, 7 $\times$
10$^{-8}$ relative to H$_2$, is reached beyond 10$^{16}$ cm,
concretely at about 4 $\times$ 10$^{16}$ cm. In this region the
formation of HNC occurs mostly through the reactions:
\begin{equation}
\rm HCNH^+ + e^- \rightarrow HNC + H \label{reac-hcnh+_e-}
\end{equation}
\begin{equation}
\rm C + NH_2 \rightarrow HNC + H \label{reac-c_nh2}
\end{equation}
where the formation of HCNH$^+$ is again driven by cosmic rays
while reaction (\ref{reac-c_nh2}) occurs as a consequence of the
photochemistry, with atomic carbon and NH$_2$ coming from the
photodissociation of acetylene and ammonia, respectively. Reaction
(\ref{reac-hcnh+_e-}) has been well studied in the laboratory
\citep{sem01}, but there are still some uncertainties
regarding the branching ratios yielding HCN and HNC. On the other
hand, the kinetics of reaction (\ref{reac-c_nh2}) is poorly known
and the rate constant has been taken from an estimate of
\citet{smi04}. The destruction of HNC in IRC~+10\,216's envelope is
dominated by photodissociation, for which the rate is taken from
udfa06$^1$. Despite the lack of accurate branching ratios and
rate constants for some of the reactions involved in 
HNC formation, its predicted abundance has the right order of magnitude,
although these inaccuracies may introduce some errors in the
abundance radial distribution calculated, as will be shown in Sect. \ref{sec:results}.

\subsection{Radiative transfer model}

The radiative transfer calculations were performed with the code
described in \cite{dan08}. The first 20 rotational levels of HNC 
in the ground and in the excited vibrational states $\nu_2=1,2$
(bending mode at 21 $\mu$m with $\ell$-type doubling), $\nu_1=1$ (NH
stretching mode at 2.7 $\mu$m), and $\nu_3=1$ (NC stretching mode at
4.9 $\mu$m) were included with the spectroscopic parameters given
by \cite{mak01}. The experimental electric dipole moment of HNC,
3.05 Debye \citep{bla76}, was used for the rotational transitions
in each vibrational state. For ro-vibrational transitions the
dipole moments were taken from \cite{har02}. The collisional rate
coefficients computed by \citet{dum10} for the first 26 rotational
levels and for temperatures between 5 and 500 K were used for pure
rotational transitions. For the densities prevailing in the region
where HNC has a significant abundance, the collisional excitation
of the vibrational levels of HNC is expected to be negligible (c.f. Sect. \ref{sec:results}). Therefore,
for ro-vibrational transitions, the same set of rate coefficients was used
but they were scaled down by an arbitrary factor of 100.

\section{Results} \label{sec:results}

The HNC line profiles that result from the abundance distribution
given by the chemical model are plotted in
Fig.~\ref{fig-hnc-lines}, as red lines. It is seen that the model
overestimates the intensity of some of the HIFI lines, in
particular the $J=6-5$. A better overall agreement is obtained if
the abundance of HNC is lowered to $7 \times 10^{-9}$, relative to
H$_2$, for $r<15\arcsec$ (see blue lines in
Fig.~\ref{fig-hnc-abundance} and Fig.~\ref{fig-hnc-lines}). It,
thus, seems that the chemical model overestimates the abundance of
HNC in the regions inward of 15\arcsec\/ by a factor of $\sim 5$.

\subsection{HNC excitation}

\begin{figure}[h!]
\begin{center}
\includegraphics[angle=0,scale=0.7]{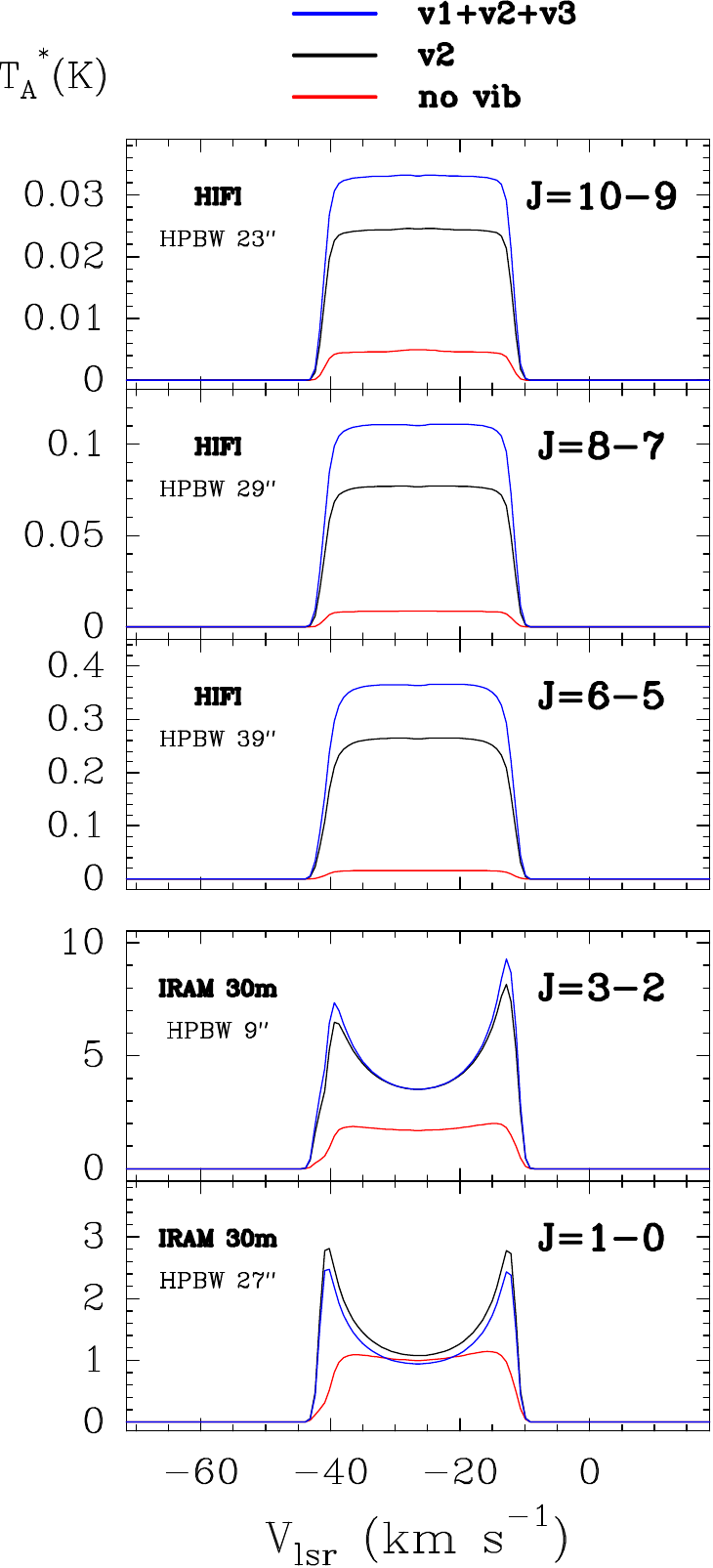}
\caption{Effect induced by the inclusion of the vibrational modes
of HNC on the lines observed by the IRAM and Herschel telescopes,
for the model with unaltered X(HNC) profile.} \label{fig-hnc-pumping} \vspace{-0.5cm}
\end{center}
\end{figure}

The influence of the different vibrational excited states on the excitation 
of the observed HNC lines is illustrated in Fig. \ref{fig-hnc-pumping}.
It is seen that the line intensities are largely increased when
the $\nu_2$=1 state is included, and to a lesser extent when the
the stretching modes are considered.
From the results shown in this figure,
it is concluded that the excitation of HNC
rotational levels is mainly dominated by infrared pumping to the
excited vibrational states, and to a lesser extent by inelastic
collisions. Therefore, results should not be substantially altered
by the uncertainty in the $T_k$ profile, something that has been
checked adopting alternative $T_k$ profiles. We note that this
also validates the use of an arbitrary set of ro-vibrational
collisional rate coefficients. This point has been checked by considering 
scaling factors that differ by a factor 2 with respect to the value of 100 used to scale 
down the ro-vibrational rate coefficients. No differences were found between the various results.
Owing to the efficiency of the radiative pumping to the 
vibrational levels, the accuracy of the radiative-transfer model 
largely relies upon the correctness of the SED. In order to test 
the error that would result from inaccuracies in the SED modeling, we adopted alternate
models for the dust composition, that result in qualitatively good representations
of the overall SED, but with variations in its shape (i.e. with maximum 
variations of the order of 10-20\% at a given frequency, in comparison 
to the current model). From these tests, it is concluded that the line fluxes will
show variations of the same order as the variations observed for 
the continuum at 11 and 21 $\mu m$. Thus, even if the quality of the 
SED can play a role in explaining the differences observed between
the observations and the model, the effect should be modest and may not
explain the differences observed (a factor $\sim$ 2 for the $J=6-5$ line).

\subsection{Density-enhanced shells}

Another aspect worth considering is the inclusion of density
enhanced shells. Such shells have been characterized 
through optical imaging by \cite{mau00} and  would originate from an episodic increase
of the mass loss rate of the star. From this study, it appears that their exact characterization is complex:
rather than continuous spherical shells, various disrupted arcs of matter are detected. 
These shells have more recently been characterized with the PACS instrument on-board Herschel \citep{decin2011}.
This study shows that the shell structure extend at least out to $\sim$320\arcsec\/ and that the shells 
have a density which is typically enhanced by a factor 4 with respect to the intershell medium. 
An attempt in introducing such shells in order to deal with the analysis of molecular 
emission has been proposed by \cite{cor09}. These authors introduce 2\arcsec\/ wide shells with an 
intershell distance of 12\arcsec\/. This model is successful in reproducing the spatial 
distribution of C$_n$H molecular species, and we therefore adopted the same structure for the shells in
the current work.
\begin{figure*}[ht]
\begin{center}
  \subfigure[PdBI observations]{\includegraphics[angle=0,scale=0.45]{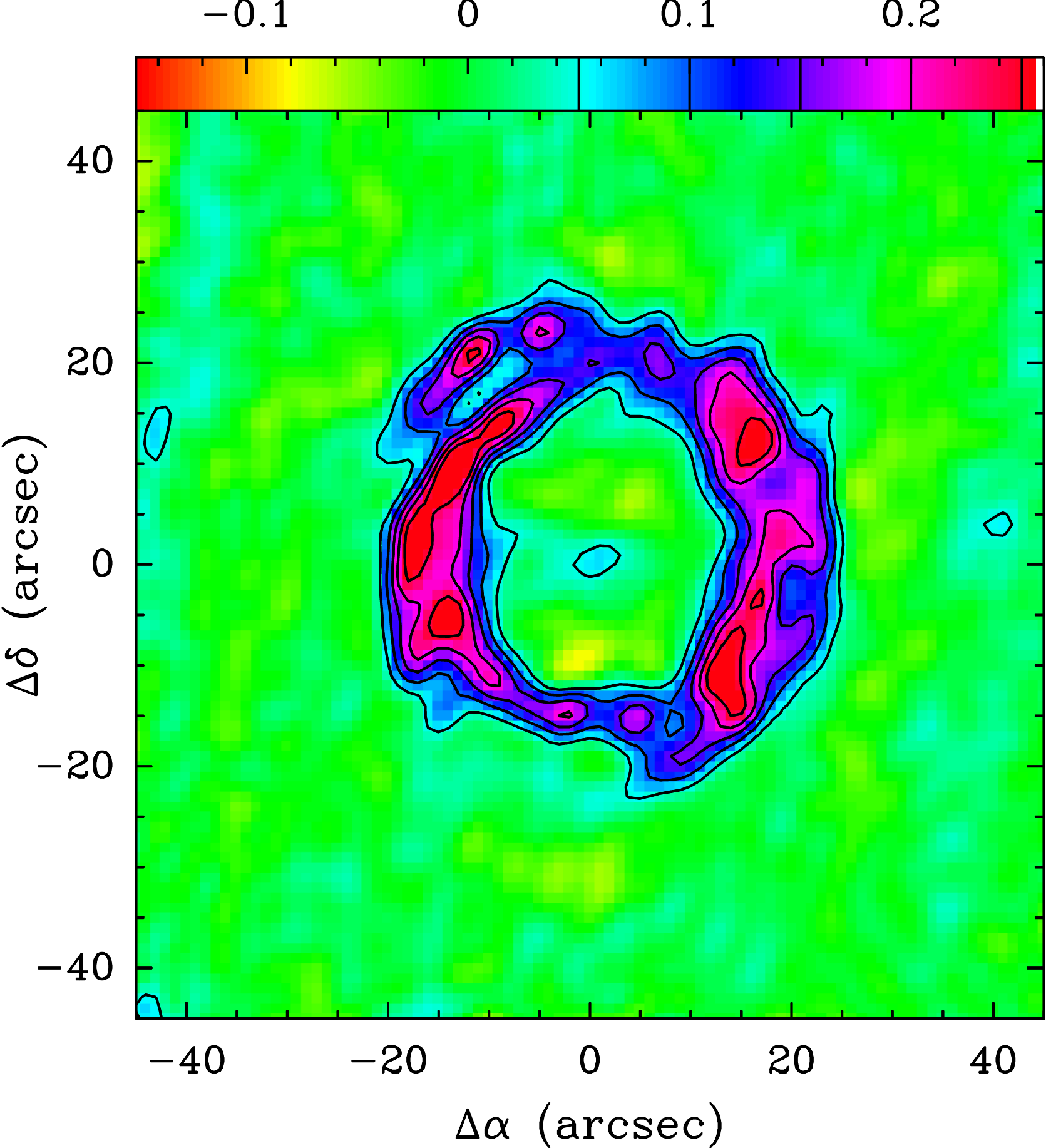}} \quad
  \subfigure[density-enhanced shells + X(HNC) central drop]{\includegraphics[angle=0,scale=0.45]{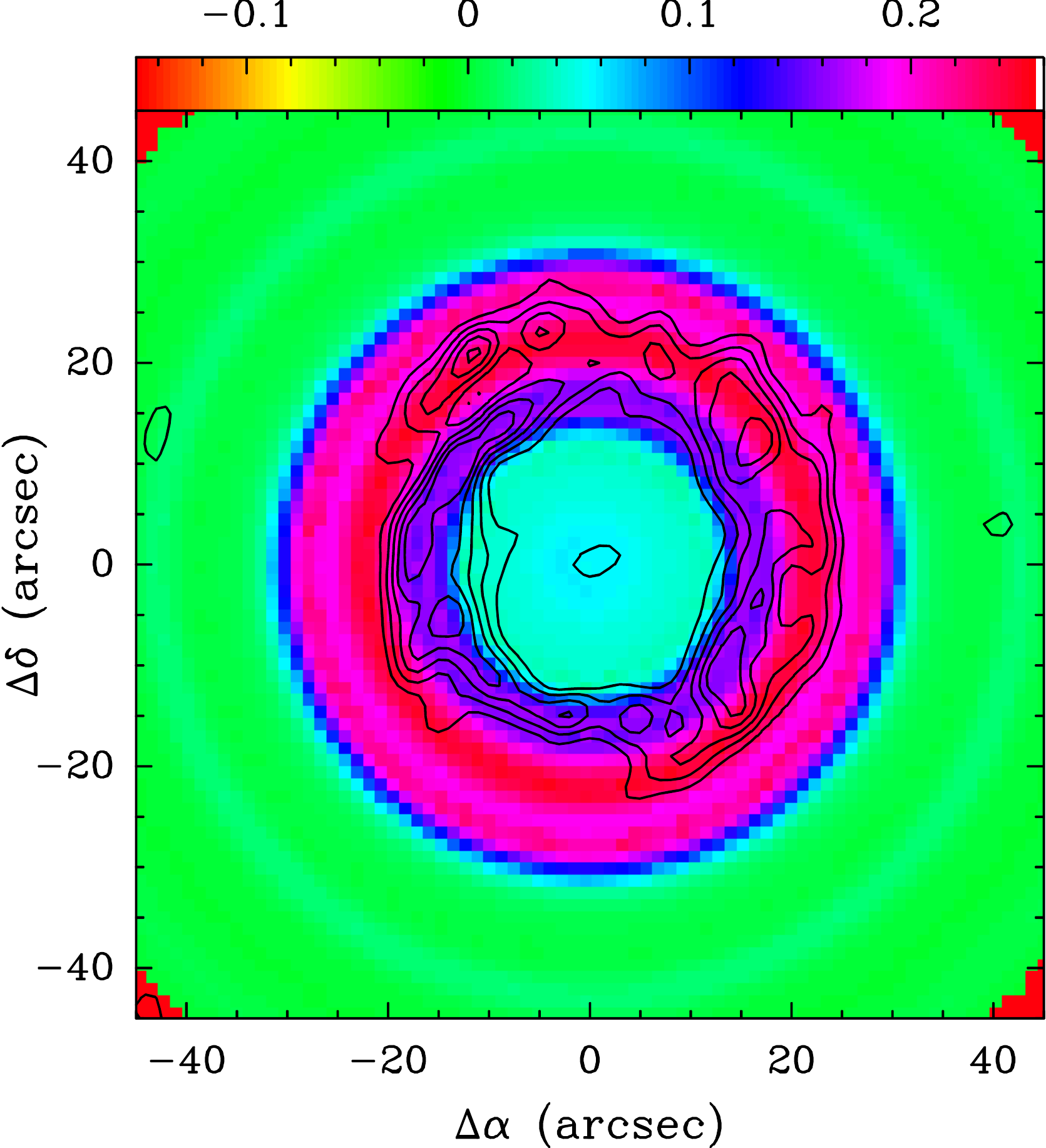}} \\
  \subfigure[density-enhanced shells]{\includegraphics[angle=0,scale=0.45]{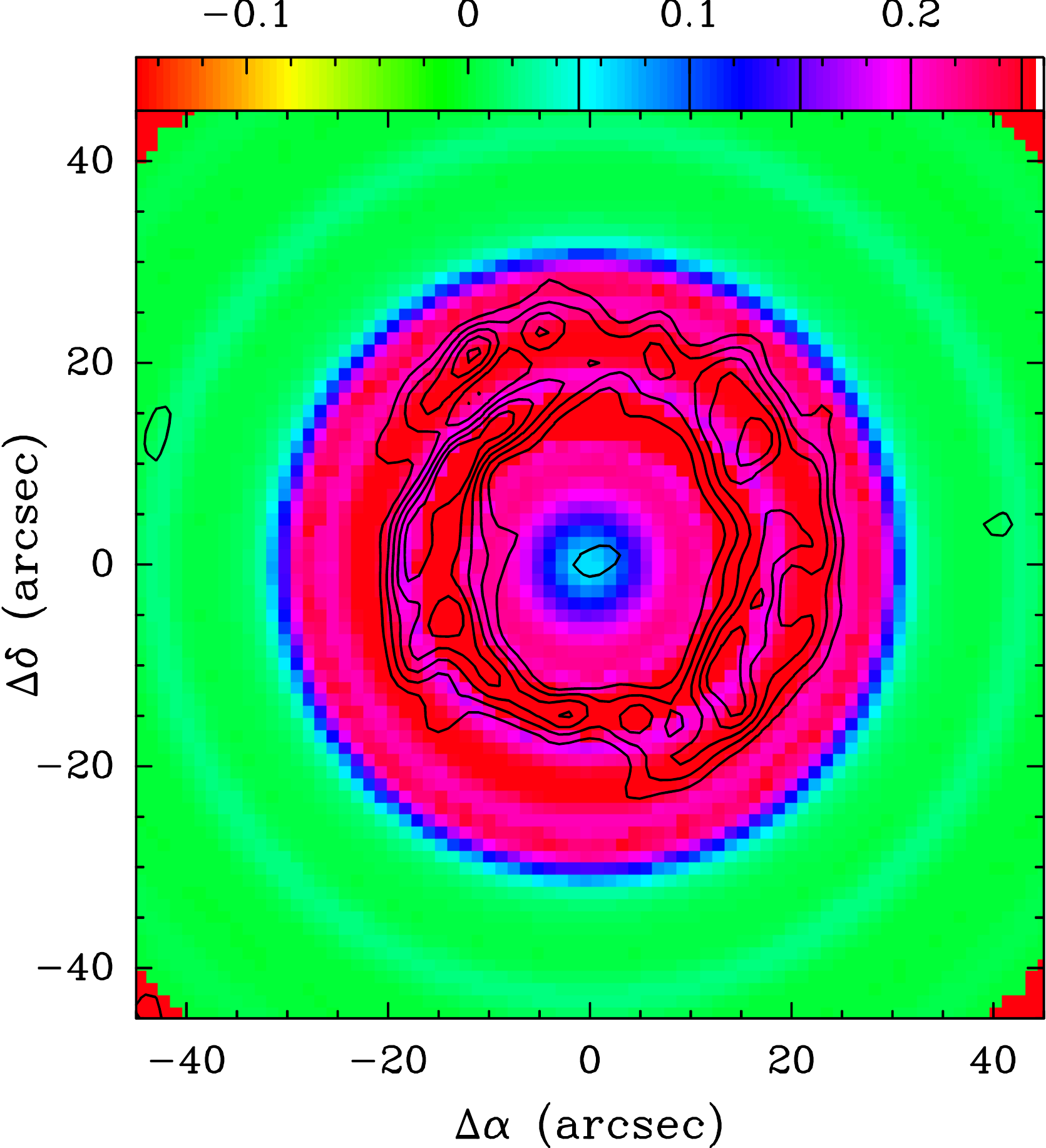}} \quad
  \subfigure[no density-enhanced shells]{\includegraphics[angle=0,scale=0.45]{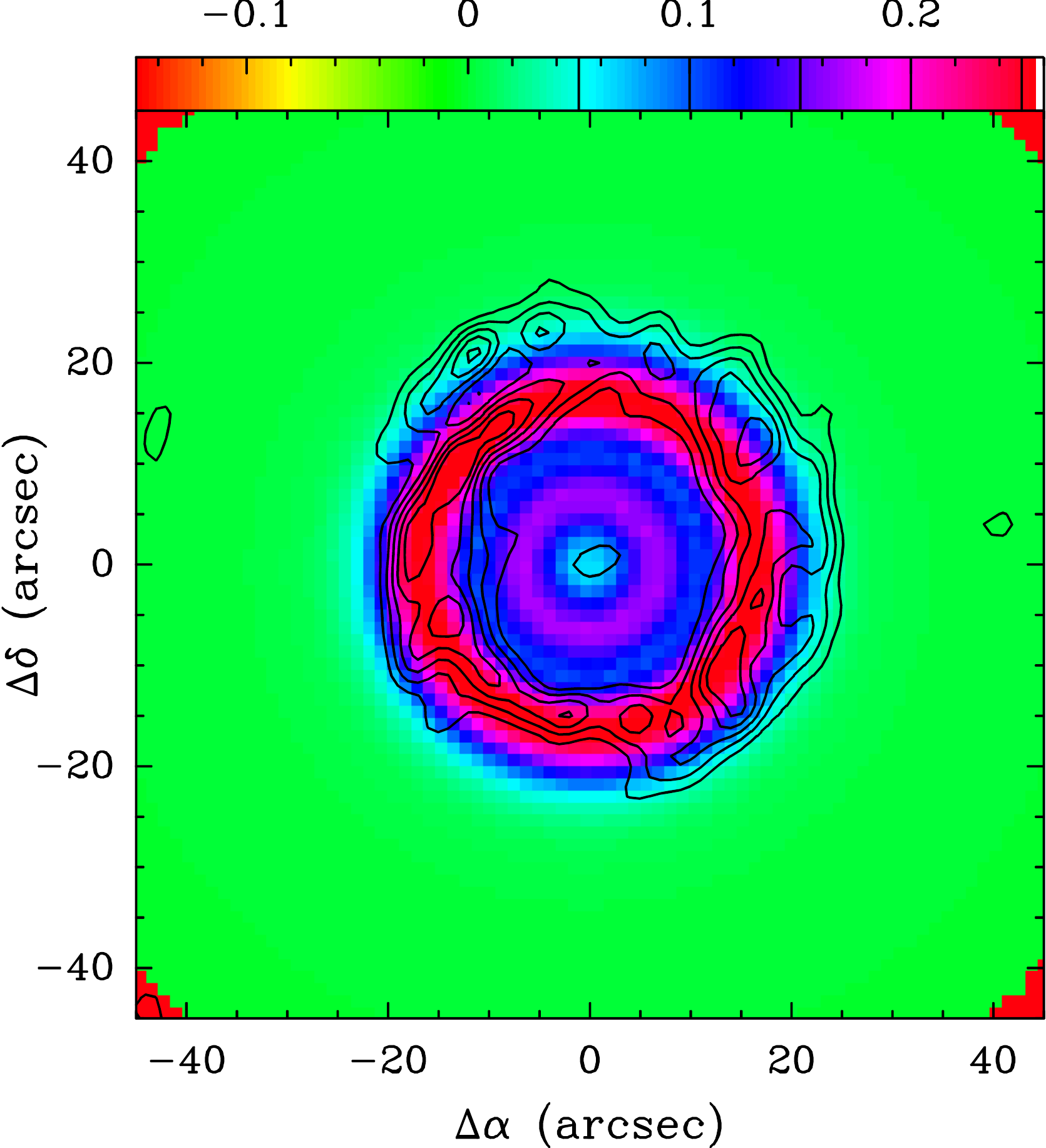}} \\
\end{center}
\caption{Images of the HNC($J=1-0$) emission integrated over the 10 km/s central velocity around the V$_{LSR}$
of the source. The fluxes are given in Jy/beam, with a beam size which correpond to the Plateau de Bure interferometer ($\sim$4\arcsec\/). 
The various panels correspond to (a) IRAM PdBI observations reported in \cite{gue97} (b)  model with density-enhanced shells
and with a central decrease of the abundance (blue curve in Fig. \ref{fig-hnc-abundance}) (c)  model with density-enhanced shells
and with the abundance given by the standard chemical model (red curve in Fig. \ref{fig-hnc-abundance}) (d)  model without density-enhanced shells and with the abundance given by the standard chemical model (blue curve in Fig. \ref{chem_ES-NS}). In panels (b), (c) and (d), the isocontours correspond to the IRAM PdBI observations displayed in panel (a).}
\label{maps}
\end{figure*}

\begin{figure}
\begin{center}
\includegraphics[angle=270,scale=0.35]{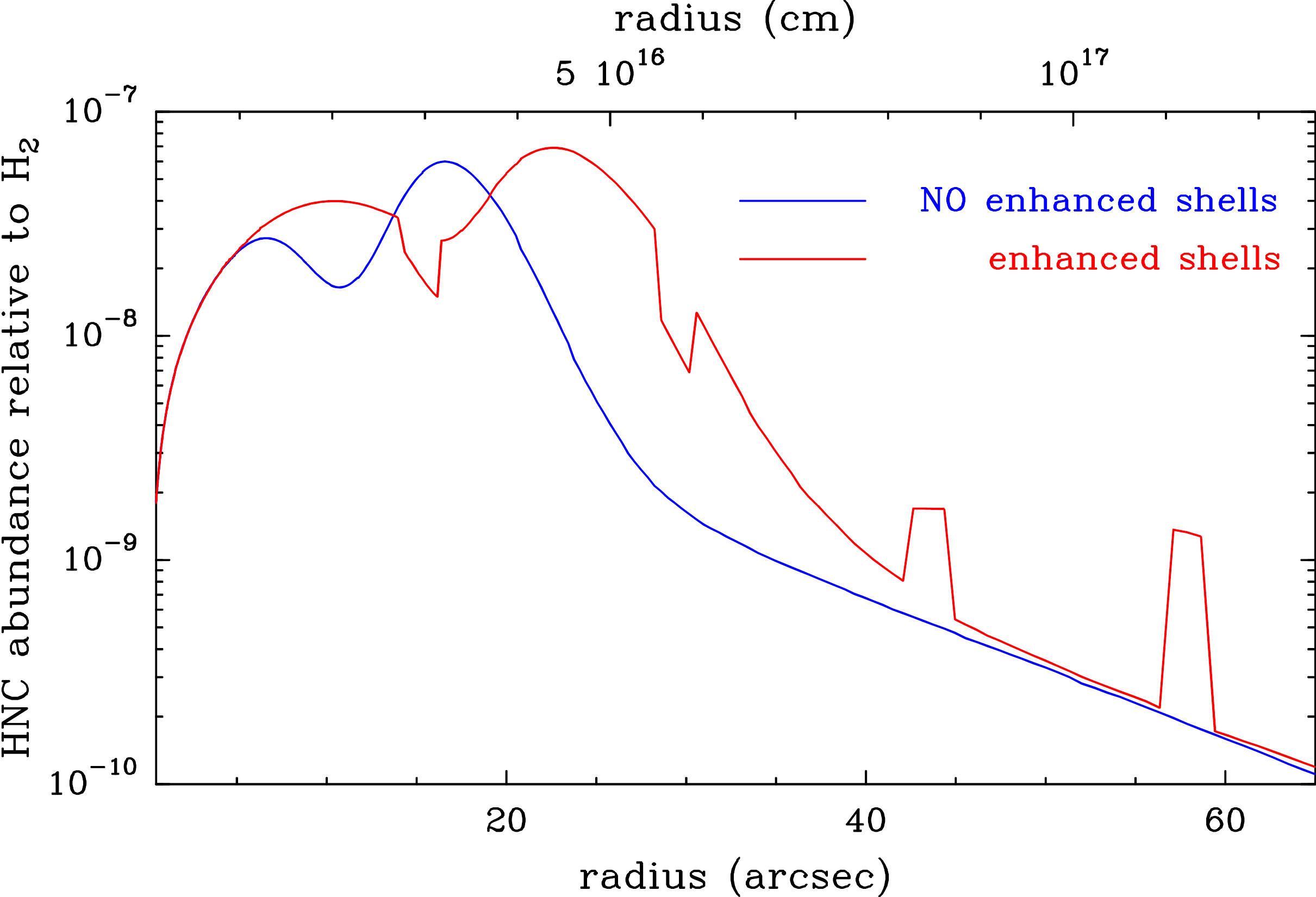}
\caption{HNC abundance as a function of radius, without (blue) and with (red) density 
enhanced shells.} \label{chem_ES-NS} \vspace{-0.5cm}
\end{center}
\end{figure}

\begin{figure}
\begin{center}
\includegraphics[angle=0,scale=0.5]{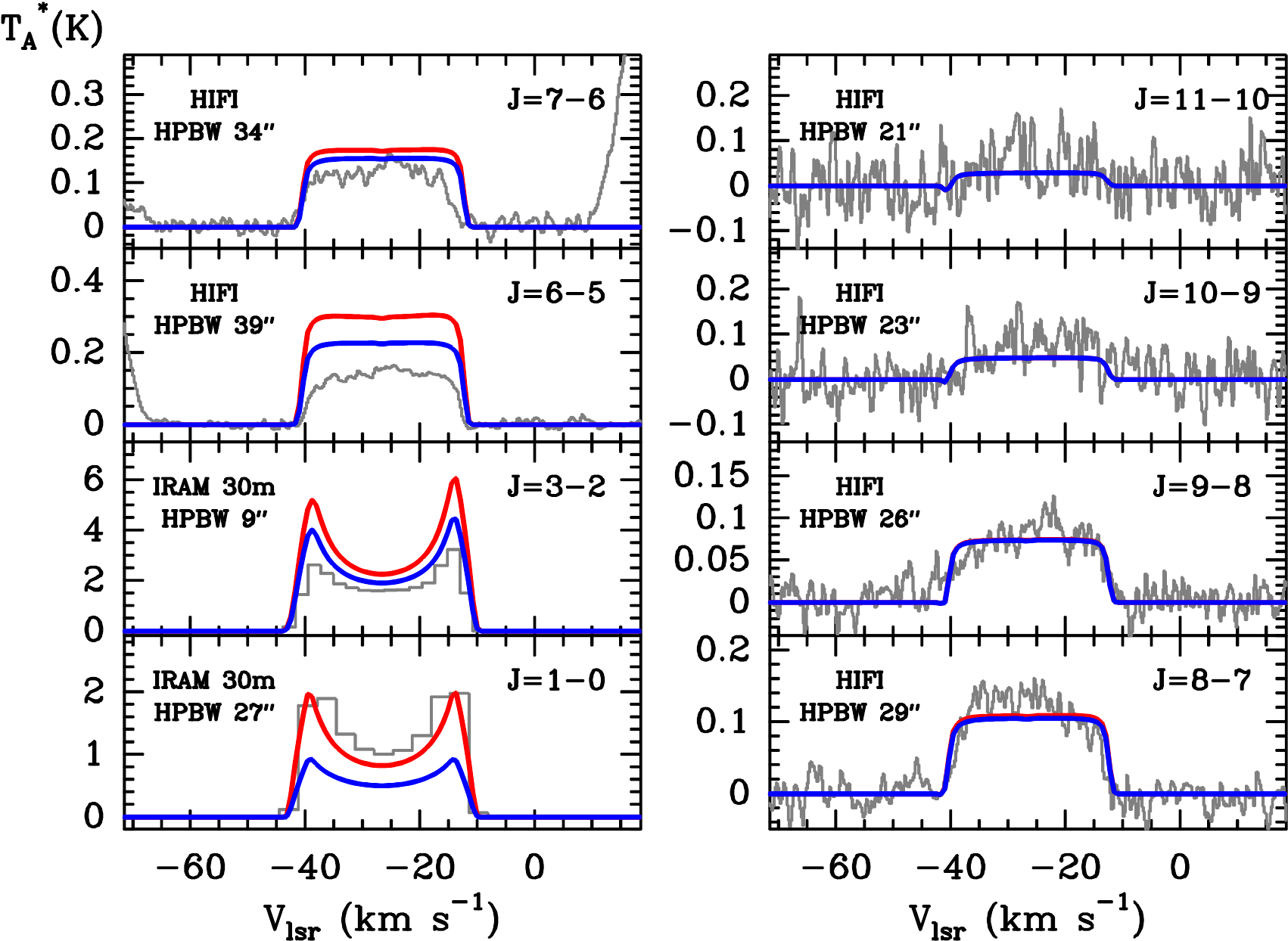}
\caption{HNC line profile for the model with density-enhanced shells (red) and for the model
without shells (blue).}\label{mod_ES-NS} \vspace{-0.5cm}
\end{center}
\end{figure}

The main effect introduced by these shells is on
the radial profile of the molecular abundance predicted by the chemical modeling. Indeed, without density-enhanced shells,
the photodissociation of the molecules is efficient down to smaller radii, as 
compared to a model including such shells, and therefore the peak abundance 
of the molecules formed by photochemistry is shifted inward.
This result is similar to the one described by \cite{debeck11} while 
dealing with the interpretation of the C$_2$H HIFI observations.
This is illustrated in Fig. \ref{chem_ES-NS} for the case of HNC
where the abundance of the molecule is presented for the two models.
The line profiles obtained for the two HNC radial distributions are shown in Fig. \ref{mod_ES-NS}.
Except for the $J=1-0$ line, omitting the density-enhanced shells would result in a better agreement
between the model and observations. Still, in the model without density-enhanced shells, the $J=6-5$ line
is overestimated, the intensity being $\sim$ 35\% higher in the model in comparison to the observation.
Fig. \ref{maps} shows the flux of the $J=1-0$ line, integrated over a 10 km/s wide
interval centered on the $V_{LSR}$ of the source. Panel (a) corresponds to the IRAM
Plateau de Bure observations described in \cite{gue97}. In this figure, the extent of the emission
predicted by the models is shown for various cases: (b) with density-enhanced shells and including a central drop
in the HNC abundance, as derived from the HIFI observations (c) with density-enhanced shells, with the abundance
profile predicted by the chemical modeling  (d) without density-enhanced shells and with the abundance
profile predicted by the chemical modeling. 
First, a comparison of panels (c) and (d) shows that the inclusion of the density-enhanced shells has a profound influence
on the extent of the emission. As commented previously, the density-enhanced shells attenuate the UV field, reducing the amount
of photodissociation. This entails that the inclusion of the density-enhanced shells increases the extent of the emission 
of the $J=1-0$ line to larger radii.
Comparing the prediction of the two models with the observations, it seems that a good fit to the observations would correspond
to a structure intermediate, in the sense of the attenuation
of the external UV field, to the models with and without density-enhanced shells considered here.
Moreover, none of these models is able to reproduce the extent of the ring-shaped structure which is observed.
On the other hand, and considering panel (b), we see that the central drop introduced to fit the HNC lines observed with HIFI
permits to reproduce this ring-shaped structure. Moreover, we note that such a decrease in the HNC abundance would have to be introduced
in any model intermediate to the two models presented in Fig. \ref{chem_ES-NS}. Indeed, the model without density-enhanced shells
corresponds to the lowest abundance possible for HNC, as predicted by the chemical modeling, for radii $r \leq 12\arcsec$. 
Whatever could be the structure considered for the shells, the abundance of HNC would have to be reduced in this region in order to reproduce 
the observed $J=6-5$ line and to fit the ring-shaped structure of the $J=1-0$ emission.
Finally, by considering various studies that deal with the geometrical
structure of IRC~+10\,216, it is evident that density-enhanced shells 
are present in the source and have to be included \citep[e.g.][]{mau00,cor09,debeck11}.
However, the model adopted in the present study is a simplification of the exact structure of the shell morphology which is observed through
optical imaging \citep{mau00} or through CO lines \citep{fong03,cer12} 
and this introduces a 
source of uncertainty in the current modeling. But, as discussed earlier when considering the ''extreme case''
of an envelope without density-enhanced shells, the fact that the HNC abundance has to be decreased with respect
to the prediction of the chemical modeling should not be altered by the description adopted to model the shells.

\subsection{HNC chemistry}

\begin{figure}
\begin{center}
\includegraphics[angle=270,scale=0.35]{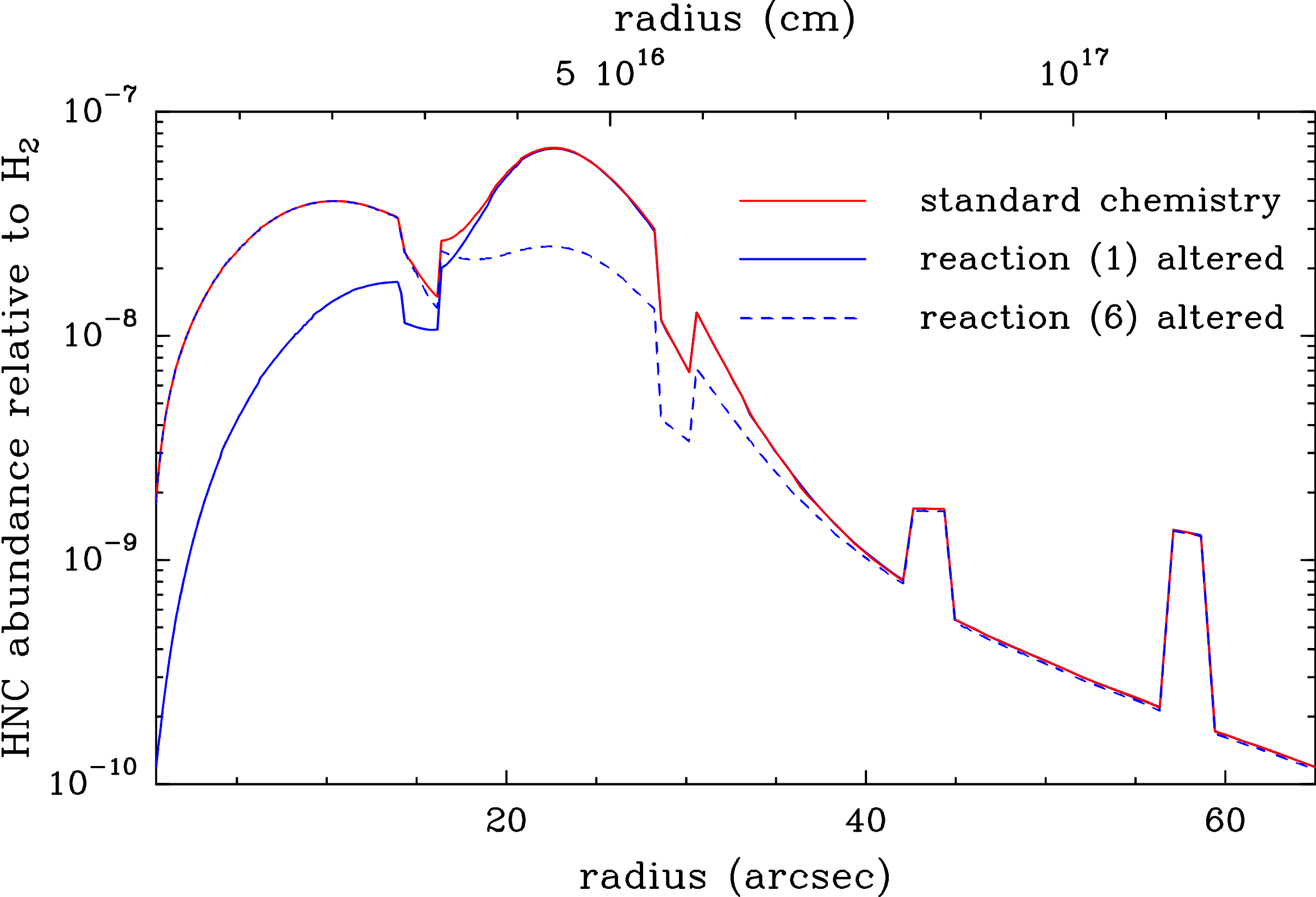}
\caption{HNC abundance profile derived for various cases. The red curve correspond to the abundance
obtained with the standard chemical network. The blue curve is obtained by changing the branching ratio of reaction (1) of Sect. \ref{section:chemistry} and the dashed blue curve removing reaction (6) from the chemical network (see text).} \label{fig8} \vspace{-0.5cm}
\end{center}
\end{figure}

\begin{figure}
\begin{center}
\includegraphics[angle=0,scale=0.5]{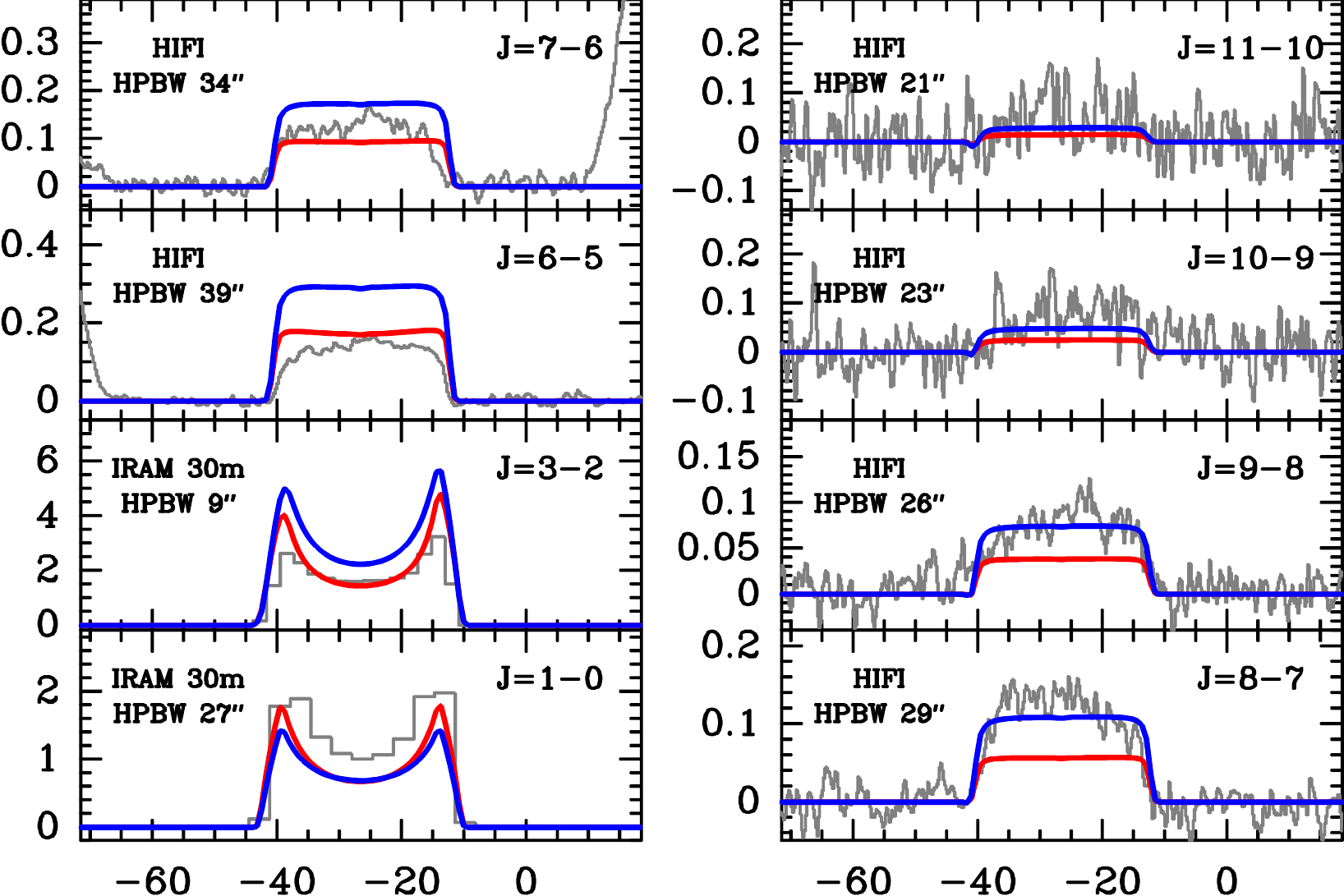}
\caption{HNC line predicted for the abundance profiles shown in Fig. \ref{fig8}. The red and blue
lines correspond respectively to an alteration of reaction (1) or (6).} \label{fig9} \vspace{-0.5cm}
\end{center}
\end{figure}

As discussed in Sect. \ref{section:chemistry}, the formation of HNC in the region $r \leq 30\arcsec$ is mainly
driven by three reactions, of which two have poorly constrained branching ratios or rate constants. In what
follows, we examine how the uncertainties on these reactions, i.e. reactions (1) and (6) of Sect. \ref{section:chemistry},
could explain why the chemical modeling overestimates the HNC abundance.
Fig. \ref{fig8} shows the HNC abundance profile obtained by modifying the rate constant of reactions (1) and (6). For the case 
of reaction (1), the alteration consists in modifying the branching ratio of the reaction and for the curve 
shown in Fig. \ref{fig8}, the abundance is shown for the case where 75\% of the collisions lead to HCN and
25\% to HNC. For the case of reaction (6),
the abundance shown in Fig. \ref{fig8} corresponds to the case where this reaction is omitted from the chemical network.
For these two cases, the corresponding line profiles are presented in Fig. \ref{fig9}.
First, we see from Fig. \ref{fig8} that altering reaction (1) can introduce a reduction of the HNC abundance
for radii below 15\arcsec\/. This is the region where the HNC abundance has been found to be overestimated when modeling 
the observations. However, considering the line profiles shown in 
Fig. \ref{fig9}, solely changing the branching ratio of this reaction would not enable to reproduce the observations. 
Indeed, fitting the $J=6-5$ line would require to lower the amount of HNC produced via this reaction below 25\% and in such a case, 
the higher excited lines would have their intensities largely underestimated. 
The removal of reaction (6) from the chemical network modifies the HNC abundance for radii $15\arcsec < r < 40\arcsec$. The effect on the 
line profile, in comparison to the standard network is rather low, and only lead to a slight lowering of the intensities of the 
$J=1-0$ and $J=3-2$ lines. 
So within to the current chemical network the most probable source of uncertainty lies in
the chemical kinetics of reaction (1) and its branching ratio. We stress that the amount of HNC produced 
via this reaction also depends on the NH$_3$ abundance. In the case of
IRC~+10\,216, the NH$_3$ abundance in the innermost part of the envelope has been derived to be in the range
$10^{-7} < $X(NH$_3$) $< 2 \times 10^{-6}$ \citep{has06,kea93}. In the current model, we assume an
abundance that corresponds to the upper limit (see Fig. \ref{fig10}). 
We tested the lower limit for the NH$_3$ central abundance and found that the decrease in the HNC abundance was lower than in the model where the branching ratio of reaction (1) is modified. 

\subsection{Conclusion}

\begin{figure}
\begin{center}
\includegraphics[angle=-90,scale=.38]{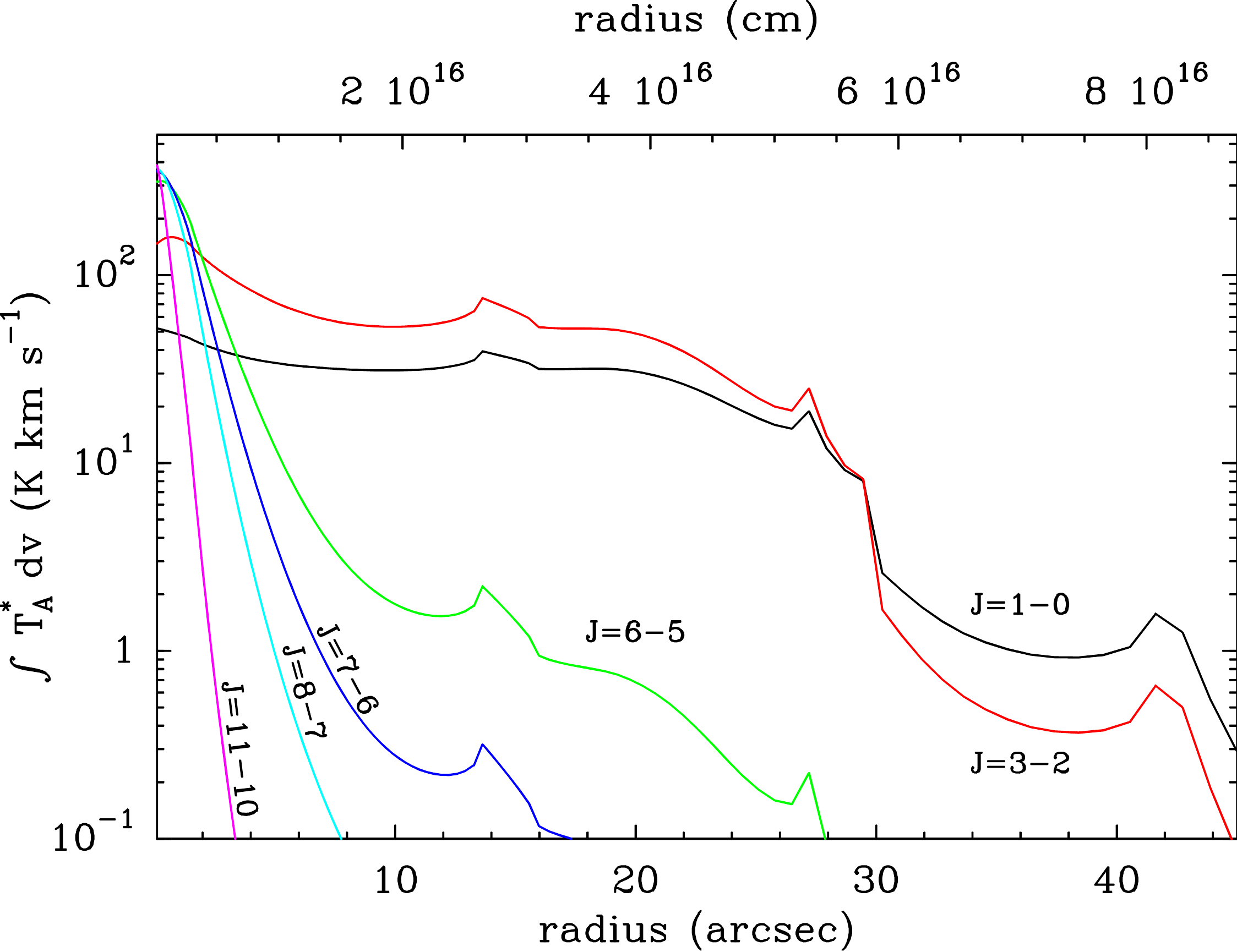}
\caption{Velocity integrated intensity as a function of impact
parameter for various HNC lines.} \label{fig-hnc-tantdistr}
\vspace{-0.5cm}
\end{center}
\end{figure}

Several conclusions may be obtained from the radiative transfer
modeling. 
Firstly, HNC transitions with higher energies have smaller emitting regions.
For instance, the emission region of the high-$J$
transitions observed with HIFI is noticeably smaller than that of
the $J$ = 1-0 transition, which emits up to a radius of 20\arcsec\/
(see \citealt{gue97}). The model clearly illustrates this on
Fig.~\ref{fig-hnc-tantdistr}, where the velocity-integrated
intensity of various lines is plotted as a function of radius. A
second conclusion is that infrared pumping to the $\nu_2$=1 state
is the main mechanism to populate the HNC rotational levels of the
ground vibrational state in IRC~+10\,216's envelope. Thus, the
emission of high-$J$ lines of HNC is favored in regions which are
warm and/or have a large radiation flux at $\lambda$ 21 $\mu$m.
It is worth here to comment on the detection of HNC $\nu_2$=1 in
the carbon-rich pre-planetary nebula CRL 618 \citep{sch03}. These
authors found that vibrationally excited HNC is confined to the
hot circumstellar layers, although it is likely pumped by
$\lambda$ 21 $\mu$m photons because of the high radiation flux at
this wavelength in CRL 618. In IRC~+10\,216, the radiative transfer
model predicts that rotational lines within the $\nu_2$=1 state
would have intensities of a few mK, and thus may have escaped
detection due to the lack of sensitivity.

\section{Summary}

The detection of highly excited rotational lines of HNC toward
IRC~+10\,216 evidences the existence of a warm HNC component whose
emission arises from regions smaller (extending up to a few
arcseconds) than those traced by the fundamental rotational
transition, previously mapped at $\lambda$ 3 mm with the PdBI, and
whose emission extends up to 20\arcsec\/ (see \citealt{gue97}). These
results are still compatible with HNC being formed not close to
the star, but in the intermediate and outer layers where 
cosmic-ray ionization drives the formation of the precursor ion HCNH$^+$.
This study has shown that the infrared pumping through the first
excited state of the bending mode $\nu_2$ dominates the excitation
of the high-$J$ rotational levels involved in the
transitions observed with HIFI. Emission in these high-$J$
rotational transitions of HNC is expected to be strong in regions
where HNC is either present in a warm and dense region and/or
surrounded by intense infrared radiation at wavelengths around 21
$\mu$m.

The present study shows that the HNC abundance can be qualitatively
reproduced by the chemical modeling and using the currently available reaction network.
However, a quantitative estimate of the HNC abundance points toward multiple origins
concerning the discrepancies between observations and the models. The discrepancies
may arise from a too simplistic description of the physical structure of the envelope and particularly
of the density-enhanced shells. Another origin could be in the description of the SED of the source or,
finally, in the uncertainties in the reaction rates, branching ratio and missing reactions in the chemical network. 
We stress that a 2D description of the chemistry and radiative transfer modeling
could enable to disentangle between these various possibilities but such a work would require
additional observations in order to constrain the geometrical structure of the source. In the next years, the observing
capabilities that will be available through the ALMA interferometer will shed light on this, since
the J=1-0, 3-2, 4-3, 7-6 and 9-8 HNC lines could then be observed in this object with high angular resolution.

\begin{acknowledgements}

HIFI has been designed and built by a consortium of institutes and
university departments from across Europe, Canada, and the United
States (NASA) under the leadership of SRON, Netherlands Institute
for Space Research, Groningen, The Netherlands, and with major
contributions from Germany, France and the US. Consortium members
are Canada: CSA, U. Waterloo; France: CESR, LAB, LERMA, IRAM;
Germany: KOSMA, MPIfR, MPS; Ireland: NUI Maynooth; Italy: ASI,
IFSI-INAF, Osservatorio Astrof\'isico di Arcetri-INAF;
Netherlands: SRON, TUD; Poland: CAMK, CBK; Spain: Observatorio
Astron\'omico Nacional (IGN), Centro de Astrobiolog\'ia
(INTA-CSIC); Sweden: Chalmers University of Technology - MC2, RSS
\& GARD, Onsala Space Observatory, Swedish National Space Board,
Stockholm University - Stockholm Observatory; Switzerland: ETH
Zurich, FHNW; USA: CalTech, JPL, NHSC. M.A. is supported by a
\textit{Marie Curie Intra-European Individual Fellowship} within
the European Community 7th Framework Programme under grant
agreement n$^{\circ}$ 235753, consolider CSD2009-00038 and
AYA2009-07304.

\end{acknowledgements}

\end{document}